\documentclass[a4paper,12pt]{article}
\usepackage[english]{babel}

\def\l{\lambda}
\def\s{\sigma}
\def\o{\omega}

\def\e{\epsilon}

\def\t{\tau}
\def\a{\alpha}
\def\b{\beta}

\def\bba{\begin{array}}
\def\eea{\end{array}}

\def\ra{\rightarrow}

\def\baselinestretch{1.2}
\def\bb{\begin{equation}}
\def\ee{\end{equation}}

\begin{document}
\large

\begin{center}

{\Large Integrable initial boundary value problems}
\end{center}
\bigskip

\begin{center}
{Ismagil Habibullin}\footnote{e-mail: habibullinismagil@gmail.com}\\

{Ufa Institute of Mathematics, Russian Academy of Science,\\
Chernyshevskii Str., 112, Ufa, 450077, Russia}\\

\end{center}
\begin{abstract}
The problem of searching boundary value problems for soliton equations consistent with the integrability property is discussed. 
A method of describing integrals of motion for the integrable initial boundary value problems for the  KP equation is suggested via Green identity.
\end{abstract}

\\
\def\baselinestretch{1.5}


\section{Introduction}

\noindent Consider the IBV problem of the general form for
the NLS equation
\begin{eqnarray}&&
iq_t=q_{xx}+c |q|^2q,\quad x>0,\,t>0,
   \label{nls}\\
&&a_1q_x+a_2q\vert_{x=0}=f(t),
\label{x=0}\\
&&q\vert_{t=0}=q_0(x),\quad q_0(x)\vert_{x\ra+\infty}\ra0,
\label{t=0}
\end{eqnarray}
which was studied by many authors. The usual scattering matrix
$s(\l,t)$ of the corresponding Dirac operator on the half-line $x>0$
depends on $t$ in a very implicit way. Namely, it satisfies the
following matrix equation
 \bb
s_t=2i\l^2[s,\s_3]+Z(q(0,t),q_x(0,t),\l)s. \label{s} \ee

Equation contains unknowns $s$, $q$ and $q_x$. How to study such kind
of equations? At the first glance it contains an extra unknown and it
is under-determined. But some implicit requirement should be valid:
$s(\l,t)$ preserves its analytical properties on the upper and lower
half planes $Im \l>0$ and $Im \l<0$. So really the equation is
correctly defined.

Different approaches to study the equation (\ref{s}) are discussed in
the literature, for instance, recently were proposed the global
relation method (Fokas) and the elimination by restriction method
(Degasperis, Manakov, Santini). The following result allows to
understand the essence of the problem (\ref{nls}), (\ref{x=0}),
(\ref{t=0}).

{\bf Theorem}, \cite{dms}. The entries $\a,\, \b$ of the scattering
matrix $s(\l,t)$ satisfy the following system of equations
\begin{eqnarray}
\a_t(k,t)=\int_{-\infty}^{\infty}{dk'\over k'-k-i0}
F_1(\a(k',t),\b(k',t),f(t))&&\label{ch1}\\
 \b_t(k,t)=\int_{-\infty}^{\infty}{dk'\over k'-k-i0}
F_2(\a(k',t),\b(k',t),f(t))&&\label{ch2}
\end{eqnarray}

Generally this system of equations with the variable coefficients is
nonlinear ($F_1,\,F_2$ -- are second degree polynomials), it is
integrable only if $f(t)\equiv0$. Thus, the IBV problem is equivalent
to the Cauchy problem for a pseudodifferential equation with two
independent variables (generally nonintegrable). Integrability is
lost when $f(t)\neq0$. Only in the homogeneous case the DMS equation
is integrable (it becomes linear). The IBV problem for $f(t)\equiv0$
is studied in details (M.Ablowitz, H.Segur 1975 when $a_1=0$ or
$a_2=0$ and R.Bikbaev, V.Tarasov, I.Habibullin 1990-91 if
$a_1a_2\neq0$). It admits soliton solutions. The asymptotics for the
large values of time are obtained for an arbitrary initial value.

If $f(t)$ is not identically zero then no hope to find exact
solutions to the IBV problem. One of the ways here is to introduce a
small parameter $0<\e\rightarrow 0$, i.e. replace $f(t)$ by $\e f(t)$
and to study the influence of the boundary by using the appropriately
developed perturbation theory.

To extract integrable cases one can apply the integrability test to
the system for the scattering matrix like (\ref{ch1})-(\ref{ch2}).
But we will testify the boundary condition using directly the Lax
pair. Suppose the equation \bb q_t=f(q,q_x,q_{xx},...) \label{eq} \ee
admits the Lax pair of the form \bb \psi_x=U(q,\l)\psi,\quad
\psi_t=V(q,q_x,...\l)\psi\label{psi}\ee Let a boundary condition of
the form \bb F(t,q,q_x,...)=0 \label{bc}\ee is imposed at the point
$x=0$. Substitute the BC (\ref{bc}) into the second equation in
(\ref{psi}): $W([q],t,\l)=V(q,q_x,...\l)|_{F(t,q,q_x,...)=0}$ and
find \bb \psi_t=W([q],t,\l)\psi\label{psix=0}\ee along the line $x=0$

The BC (\ref{bc}) is consistent with the Lax pair (\ref{psi}) if
the linear equation (\ref{psix=0}) admits an additional discrete
symmetry such that there exists a matrix valued function
$H([q],t,\l)$ and an involution $h=h(\l)$ such that the
transformation $\psi\rightarrow\bar{\psi}=H\psi$ converts a
solution $\psi$ of the equation (\ref{psix=0}) into a solution.
In terms of the potentials this requirement reads as \bb
H_t(\l)=W(\l)H(\l)-H(\l)W(h(\l))\label{Ht}\ee

{\bf Example 1}, see \cite{h-v}. Consider the Korteweg-de Vries
equation $u_t=u_{xxx}-6uu_x$. The coefficient matrices for the Lax
pair are defined as $$ U=\pmatrix{0&1\cr u-\l&0},$$$$
V=\pmatrix{u_x&-4\l-2u\cr u_{xx}-(4\l+2u)(u-\l)&-u_x}.$$

Suppose the BC imposed at $x=0$ is of the form $$u=F_1(u_x,t),\quad
u_{xx}=F_2(u_x,t).$$ To look for the discrete symmetry we must solve
the equation $$ {dH\over dt}=\pmatrix{F_1&-4\l-2u\cr
F_{2}-(4\l+2u)(u-\l)&-F_1}H-\hspace{3cm}\qquad$$$$
\quad-H\pmatrix{F_1&-4h(\l)-2u\cr F_{2}-(4h(\l)+2u)(u-h(\l))&-F_1}$$

Here unknowns $F_1$, $F_2$, $H=H(u,u_x,u_{xx},...),$ $h=h(\l)$ are
uniquely found. The answer is $$ H=\pmatrix{2\l+a&0\cr
0&a-\l+\sqrt{3a^2-b-3\l^2}}, $$ $$
h(\l)={-\l+\sqrt{3a^2-b-3\l^2}\over2}.$$ The BC is of the form $$
u|_{x=0}=a,\quad u_{xx}|_{x=0}=b,$$ where $a$ and $b$ are arbitrary
constants.

{\bf Example 2}, \cite{h-v}. The Harry Dym equation
$u_t+u^3u_{xxx}=0$ admits two kinds of BC:

i) $u|_{x=0}=0,\quad  u_{x}|_{x=0}=b,$ $$H=\pmatrix{1&0\cr e^{4\l
bt}&1},\quad h(\l)=\l;$$

ii) $u_x|_{x=0}=au,\quad u_{xx}|_{x=0}=a^2u/2+b/u,$
$$H=\pmatrix{\l&0\cr (\l-h(\l))a/2&h(\l)},$$$$
h(\l)={-b-2\l+\sqrt{b^2-4b\l-12\l^2}\over4};$$ where $a$, $b$ are
constants.

{\bf Example 3}, \cite{h-k}. The discrete Heisenberg model
\begin{eqnarray}
&& (T_m-1)\frac{1}{q-q_{-1,0}}=(T_n-1)\frac{1}{q-q_{0,-1}},
\label{l1}
\end{eqnarray}
has the following Lax pair $$ L=\left(\begin{array}{cc} \lambda
-\frac{q_{-1,0}}{q-q_{-1,0}} & -\frac{qq_{-1,0}}{q-q_{-1,0}}
\\ \frac{1}{q-q_{-1,0}} & \lambda +\frac{q}{q-q_{-1,0}}
\end {array}\right),$$$$
A=\left(\begin{array}{cc} \lambda -\frac{q_{0,-1}}{q-q_{0,-1}} &
-\frac{qq_{0,-1}}{q-q_{0,-1}}
\\ \frac{1}{q-q_{0,-1}} & \lambda +\frac{q}{q-q_{0,-1}}
\end {array}\right).$$
In this case the discrete involution and the cutting off condition
are found from the equation \bb
H(m+1,\lambda)A(m,N,\lambda)=A(m,N,h(\lambda))H(m,\lambda).
\label{meq} \ee The BC reads as \bb q_{m,0}=\frac
{cq_{m,1}+(-1)^ma}{c+(-1)^mbq_{m,1}},\label{bc1}\ee where $a,b,c$ are
arbitrary constants and $a^2+b^2\neq0$. The matrix $H$ takes the form
$$H(m,\lambda)=\left(\begin{array}{cc} 1 & (-1)^mac(2\lambda+1) \\
(-1)^mbc(2\lambda+1) & 1
\end {array}\right),$$ and the involution is $\quad h(\lambda)=-\lambda-1.$

\section{How to use the discrete symmetry in the ISM}

Let us discuss how to use the discrete symmetry when constructing
solutions of the corresponding IBV problems. Take the IBV problem for
the KdV equation with vanishing BC (see, \cite{h})
\begin{eqnarray}&& u_t=u_{xxx}-6uu_x,\quad x>0,\,\,t>0,
                          \label{kdv}\\
&&u\vert_{x=0}=0,\quad u_{xx}\vert_{x=0}=0,
                          \label{kx=0}\\
&&u\vert_{t=0}=u_0(x),\quad u_0(x)\vert_{x\ra+\infty}\ra0.
                          \label{kt=0}
\end{eqnarray}
In this case $H$ is the unity matrix and
$h(\l)=\l{-1+i\sqrt{3}\over2}=\l\omega .$ Actually, the discrete
symmetry reflects only the fact that the evolution of the scattering
matrix \bb s_t=4i\l^3[s,\s_3]+u_x(0,t)\s_1s \label{1s} \ee is
invariant under the change $\l\rightarrow \o\l.$ This equation is
really nonlinear but the discrete symmetry allows one to linearize
it. Put $z=\l^3$ and define the matrices
\begin{eqnarray}
&&c_+(z,t)=(s_1(\o\l,t),s_2(\l,t))\nonumber\\
&&c_-(z,t)=\s_1\bar c_+(\bar z,t)\s_1\nonumber
\end{eqnarray}
These two matrices satisfy the Riemann problem \bb
c_+(z,t)=c_-(z,t)p(z,t),\label{rp}\ee where $p(z,t)=
e^{-4iz\s_3t}p(z,0)e^{4iz\s_3t}.$ Now it is easy to see that the
scattering matrix $s(\l,t)$ is found from the linear equation
(\ref{rp}). Using this fact one can prove

{\bf Theorem}.  Let the initial value satisfy the conditions

1) $u(x,0)=u_0(x)$ is smooth and vanishes;

2) the associated Sturm-Liouville operator has no discrete
eigenvalues,

3) the scattering matrix is unbounded at $\l=0$

\noindent then the problem (\ref{kdv}), (\ref{kx=0}), (\ref{kt=0}) is
uniquely solvable for all $t>0$. The function $u_x(0,t)$ satisfies
the following representation $$u_x(0,t)={1\over t}+o({1\over
t}),\quad t\rightarrow\infty$$

In this case two of three functions $u(0,t)$, $u_x(0,t),$
$u_{xx}(0,t)$ are zero and the third one slowly decays. It is not
ever in $L_1,$ only in $L_2.$

Unfortunately, there is no regular soliton-like solutions of the KdV
equation with the vanishing boundary conditions. If the parameters
$a$ and $b$ are different from zero then regular exact soliton-like
solutions (as well as finite-gap solutions) exist approaching
$C=\sqrt{a^2-b/3}$ at $x=\infty$ and satisfying the BC at $x=0$. They
are described in \cite{Adler}. In this case
time evolution of the scattering matrix is reduced to a Riemann
problem on a Riemann surface defined by the function $
h(\l)={-\l+\sqrt{3a^2-b-3\l^2}\over2}.$

\section{Discrete symmetry and BC in multidimensional case}

Consider the well known 2D-Toda chain
\begin{equation}
\label{toda} u_{xt} \left( {n} \right) = \exp\{ u\left( {n - 1}
\right) - u\left( {n} \right)\} - \exp\{ u\left( {n} \right) -
u\left( {n + 1} \right)\} ,
\end{equation}
with the following Lax pair
\begin{eqnarray}
&\label{todan} \phi \left( {n + 1} \right) = \left( {D_{x} + u_{x}
\left( {n} \right)} \right)\phi \left( {n} \right),\\
&\label{todaxt} \phi _{xt} \left( {n} \right) =- u_{x} \left( {n}
\right)\phi _{t} \left( {n} \right) - \exp\{ u\left( {n - 1} \right)
- u\left( {n} \right)\} \phi \left( {n} \right).
\end{eqnarray}
Impose a cutting off constraint at $n=0$ \bb\label{tdbc} f(u(-1),
u(0))=0.\ee How to find all integrable cases only by using the
equation (\ref{todaxt})? To this end it is necessary to study the
discrete symmetries of the equation (\ref{todaxt}), appearing under
the BC. But now there is no $\l$ and we are to find some
generalization of the involution $\l\rightarrow h(\l)$. It is evident
that the Toda chain is invariant under transform $x\leftrightarrow
t$, so the following pair of equation
\begin{eqnarray}
&\label{todan2} \psi \left( {n + 1} \right) = \left( {D_{t} + u_{t}
\left( {n} \right)} \right)\psi \left( {n} \right),
\\
&\label{todaxt2} \psi _{xt} \left( {n} \right)=- u_{t} \left( {n}
\right)\psi _{x} \left( {n} \right) - \exp\{ u\left( {n - 1} \right)
- u\left( {n} \right)\} \psi \left( {n} \right).
\end{eqnarray}
gives also a Lax pair to the Toda chain.

{\bf Proposition}, \cite{h-g}. Suppose that there exists such an
operator $M=aD^2_x+bD_x+c$ that for $n=0$ for any solution $\psi$ of
the equation (\ref{todaxt2}) the function $\phi=M\psi$ is a solution
of (\ref{todaxt}). Then the BC (\ref{tdbc}) takes one of the forms
below
\begin{eqnarray}
&1)& e^{u(-1)}=0, \nonumber \\ &2)& u(-1)=0, \nonumber\\
&3)&u(-1)=u(0), \label{bct}\\&4)&u_x(-1)=-u_t(0)e^{-u(0)-u(-1)}.
\nonumber
\end{eqnarray}
The corresponding operator $M$ is respectively of the form
\begin{eqnarray}
&1)& M_1=a_0e^{u}D^2_x+(b_0e^ {u}+a_0u_xe^ {u})D_x, \nonumber
\\ &2)& M_2=e^ {u}D^2_x+u_xe^ {u}D_x, \nonumber\\
&3)&M_3=e^ {u}D_x, \label{ot}\\
&4)&M_4=e^ {u}D^2_x+u_xe^ {u}D_x+e^ {-u}. \nonumber
\end{eqnarray}
where $a_0$, $b_0$ -- arbitrary constant parameters, and $u=u(0)$.
Notice that the operator $M_1$ is a linear combination of the
operators $M_2$ and $M_3$.

All of the cutting off conditions above are known to be consistent
with the integrability. Initiated by this example we formulate the
discrete involution test for multidimensional equations. Two Lax
pairs which are not connected by any conjugation transformation,
should become conjugate after imposing the BC.

Apply now the test to look for boundary conditions to the KP
equation
\begin{eqnarray}\label{kp}
v_{\tau}  + v_{xxx} - 6vv_{x}& =& 3w_{y},\\
w_x&=&v_{y},\nonumber
\end{eqnarray}
admitting the Lax pair
\begin{eqnarray}\label{xy}
&&\phi _{xx} = i\phi _{y} + v\phi,\\
&&\label{xtau} \phi _{\tau}=-4\phi _{xxx} + 6v\phi _{x}
+3(v_x+iw)\phi.
\end{eqnarray}
The equation (\ref{kp}) is invariant under the change $y\rightarrow
-y$, $w\rightarrow -w$ and by this reason the following system of
equations is also a Lax pair for the KP
\begin{eqnarray}\label{xy2}
&&\psi _{xx} =-i\psi _{y} + v\psi,\\
&&\label{xtau2} \psi _{\tau}=-4\psi _{xxx} + 6v\psi _{x}
+3(v_x-iw)\psi.
\end{eqnarray}

{\bf Proposition}, \cite{h-g}. Suppose that there exists a
differential operator $M=aD^2_x+bD_x+c$ such that for $y=0$ for any
solution $\psi$ of the equation (\ref{xtau}) the function defined as
$\phi=M\psi$ is a solution to (\ref{xtau2}). Then one of the
following equations holds
\begin{eqnarray}
&1)& w|_{y=0}=0, \nonumber
\\ &2)& (v_x-iw)|_{y=0}=0,\label{bckp}
\\ &3)& (w_{\tau}-2v_{xxy}+6iv_{yyx}+ 6v_xw-6vw_x-\nonumber\\
&&-6iw^2-12cv_y)|_{y=0}=0,\nonumber
\end{eqnarray}
where $c=c(x,\tau)$ is a solution of the equation
$c_x=(-v_x+{i\over2}w)|_{y=0}$. The corresponding operator $M$ is
of the form
\begin{eqnarray}
&1)& M=1, \nonumber
\\ &2)& M=D_x, \label{okp}
\\&3)& M=D_x^2+c.\nonumber
\end{eqnarray}

If one replaces $\psi\leftrightarrow\phi$ one gets one more
constraint \bb4)\quad
(v_x+iw)|_{y=0}=0\hspace{4cm}\label{bckp4}\ee

\section{Integrals of motion}

To look for integrals of motion we will use the Green identities
which are in this case as follows
\begin{equation}\label{gr1}
{d\over dx}(\phi_x\psi-\phi\psi_x)=i{d\over dy}(\phi\psi)
\end{equation}
and
\begin{equation}\label{gr2}
{d\over d\tau}(\phi_x\psi-\phi\psi_x)=4{d\over dy}(\psi\phi_y-
\psi_y\phi -{i\over2}v\phi\psi+i\phi_x\psi_x).
\end{equation}
Suppose that eigenfunctions $\psi$ and $\phi$ are defined as
follows
\begin{equation}\label{asym}
\phi(x,y,\tau,k)=e^{-ik^2y+kx-4k^3\tau}(1+\sum^{\infty}_{j=1}
k^{-j}\phi_j),
\end{equation}
\begin{equation}\label{asym2}
\psi(x,y,\tau,k)=e^{ik^2y-kx+4k^3\tau}(1+\sum^{\infty}_{j=1}
k^{-j}\psi_j)
\end{equation}
for $k\rightarrow\infty$, and satisfy the asymptotic requirements
 $$ \phi e^{ik^2y-kx+4k^3\tau}, \psi e^{-ik^2y+kx-4k^3\tau}
 \rightarrow1$$ for $x\rightarrow-\infty$ and for
$x\rightarrow+\infty$, respectively, then the function $F(k)$
\begin{equation}\label{gf}
  F(k)=\int^{\infty}_{-\infty}(\phi_{x}\psi-\phi\psi_x-2k)dy
\end{equation}
is a generating function of the conserved quantities. Actually, by
using the Green identities (\ref{gr1}), (\ref{gr2}) one gets $
\frac{\partial}{\partial \tau }F(k)= \int^{\infty}_{-\infty}
\frac{\partial}{\partial \tau} (\phi_{x}\psi-\phi\psi_x-2k)dy=
\int^{\infty}_{-\infty}{d\over dy}(\phi\psi_y- \phi_y\psi
+{i\over2}v\phi\psi-i\phi_x\psi_x) dy=0$.

Equations (\ref{xy})-(\ref{xtau2}) admits one more Green identity
\begin{equation}\label{gr3}
 i{\partial\over\partial\tau}(\phi\psi)=4{\partial\over\partial x}
(\psi\phi_y-\phi\psi_y-{i\over2}v\phi\psi+i\phi_x\psi_x),
\end{equation}
which allows one to find the generating function of integrals of
motion for the initial boundary value problem on the half-plane
\begin{equation}\label{f1}
F_1(k)=\int^{\infty}_{0}(\phi_{x}\psi-\phi\psi_x-2k)dy
+i\int^{x}_{-\infty}(\phi\psi-C(k))|_{y=0}ds,
\end{equation}
Here the integrand in the first integral is taken at $(x,y,\t,k)$,
and in the second -- at $(s,0,\t,k)$. Really, by means of the
identities (\ref{gr1}), (\ref{gr2}), (\ref{gr3}) one gets ${\partial
F_1\over\partial\tau}=0$. Taking the first coefficients gives

{\bf Proposition}. The KP equation on the half-plane
$y>0$, ${-\infty}<x< {\infty}$ with any of BC (\ref{bckp}.1),
(\ref{bckp}.2), (\ref{bckp4}) preserves the energy
$$J_2=\int^{\infty}_{0}\int^{\infty}_{-\infty}v^2(x,y)dxdy=const $$

\end{document}